\documentclass[11pt]{article}
\usepackage{amsfonts, graphicx, amsthm, amsmath, bm, hyperref, graphics, enumerate, color}
\usepackage[portuges]{babel}

\usepackage[portuges]{babel}
\def\ii{\'{\i}}

\begin{document}
\title{Medida de $g$ com a placa Arduino em um experimento simples de queda-livre}
\author{H Cordova \footnote{e-mail: hercilioc@hotmail.com} $\&$  A C Tort \footnote{e-mail: tort@if.ufrj.br.}\\
Mestrado Profissional em Ensino de F\'{\i}sica -- Instituto de F\'{\i}sica
\\
Universidade Federal do Rio de Janeiro\\
Caixa Postal 68.528; CEP 21941-972 Rio de Janeiro, Brazil}

\maketitle
\begin{abstract}
Um modo simples de medir a acelera\c c\~ao da gravidade $g$ no laborat\'orio de f\ii sica do ensino m\'edio e de f\ii sica b\'asica no ensino universit\'ario com um microcontrolador da fam\ii lia  Ardu\'{\i}no \'e proposto. Resultados experimentais com um erro relativo de $0,1\,\%$ s\~ao apresentados e comparados com o valor local de $g$ medido pelo Observat\'orio Nacional, Rio de Janeiro.    
\bigskip
\end{abstract}
Palavras-chave: acelera\c c\~ao da gravidade, queda-livre, Arduino, 
\newpage
\section{Introdu\c c\~ao}
Um dos muitos modos poss\ii veis de medir a acelera\c c\~ao da gravidade $g$ \'e deixar cair verticalmente um corpo a partir de uma altura pr\'e-determinada $h$ e medir a dura\c c\~ao do seu tempo de queda. Como a velocidade inicial \'e nula, segue que:

\begin{equation}
g= \frac{2h}{t^2},
\end{equation}
onde $t$ \'e a dura\c c\~ao da queda. Embora a Eq. (1) possa ser introduzida nos primeiros est\'agios do ensino da cinem\'atica e seja o fundamento de uma das maneiras mais simples de medir $g$, a rapidez com que a queda-livre acontece pode tornar o experimento frustante para um iniciante. \'E poss\ii vel melhorar os resultados por meio de circuitos eletr\^onicos capazes de determinar intervalos de tempo com pelo menos $1/10$ de milissegundo de resolu\c c\~ao, veja por exemplo \cite{Blackburn&Koenig1976}, mas estas t\'ecnicas s\~ao mais apropriadas para os laborat\'orios did\'aticos avan\c cados dos cursos de gradua\c c\~ao. Por outro lado, o uso cada vez mais difundido dos modernos microcontroladores de baixo custo, como por exemplo, a fam\ii lia de microcontroladores ou placas Ardu\ii no  \cite{Cavalcanteetall2008, Cavalcanteetall2011, Salim2011}, capazes de medir intervalos de tempo na faixa de mili e microssegundos permite obter resultados perfeitamente aceit\'aveis no laborat\'orio de f\ii sica do ensino m\'edio e de f\ii sica b\'asica no ensino universit\'ario, mesmo que em uma primeira abordagem, por conveni\^encia pedag\'ogica, desprezemos os efeitos da resist\^encia do ar e outros efeitos esp\'urios.  Nas pr\'oximas se\c c\~oes discutiremos como isto pode ser feito.   
\section{Arranjo experimental}
\begin{figure}[t!]
\centering
\includegraphics[scale=0.450]{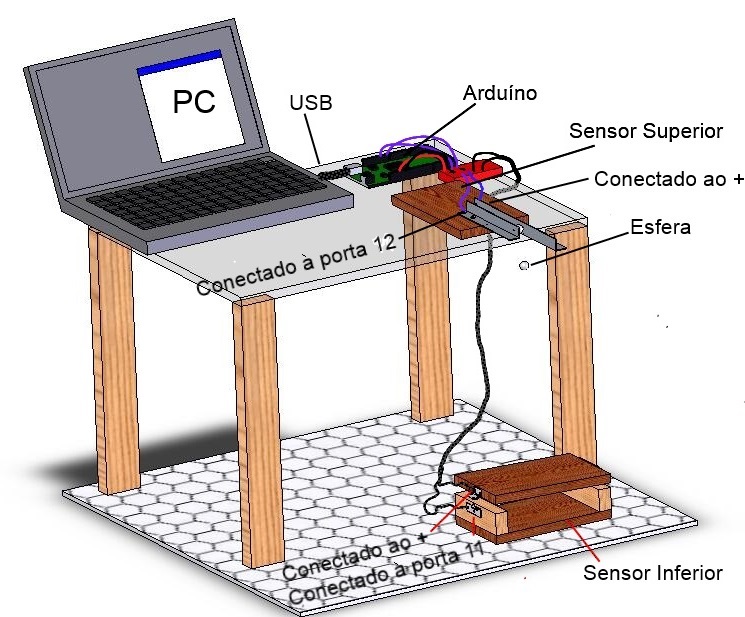}
\caption{Vis\~ao geral do arranjo experimental. }
\label{esquema}
\end{figure}

\begin{figure}[b!]
\centering
\includegraphics[scale=0.30]{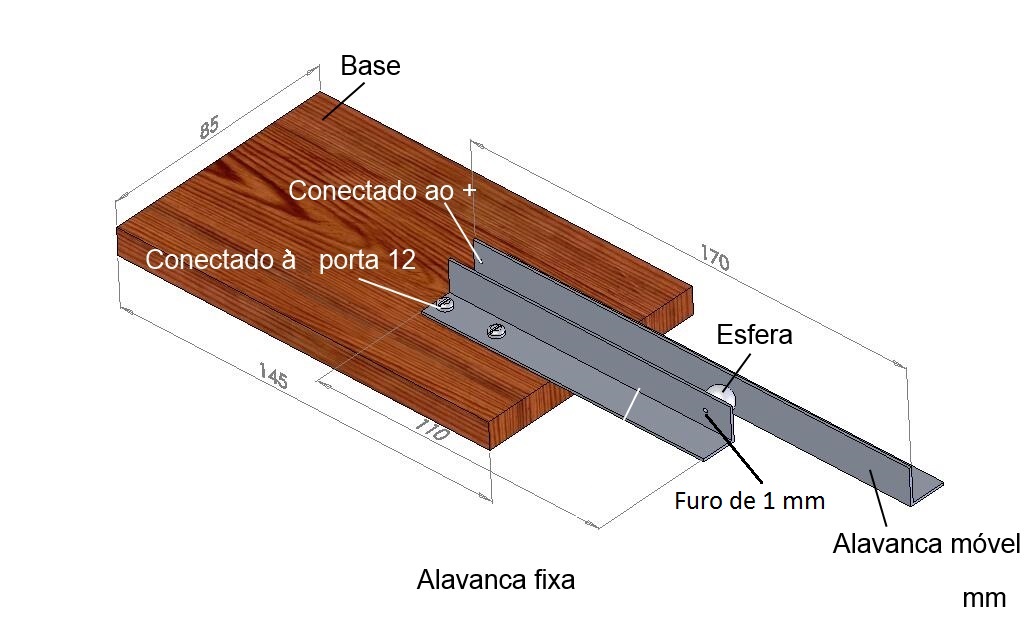}
\includegraphics[scale=0.30]{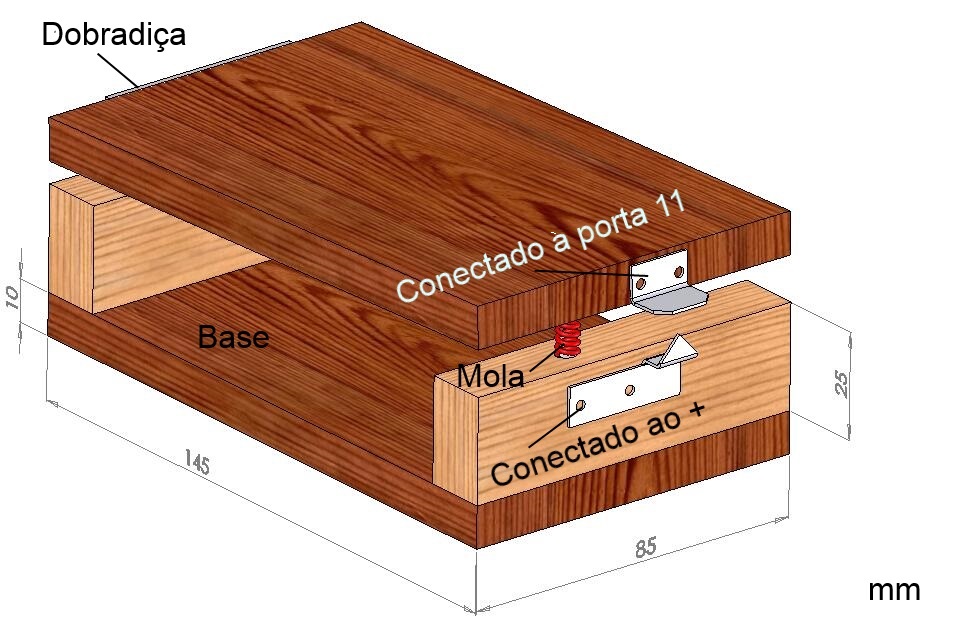}
\caption{Sensores superior (esquerda) e inferior (direita). No desenho, por clareza, o espa\c co entre os contatos el\'etricos no sensor inferior est\'a exagerado, no experimento ele \'e inferior a $1\,$mm.}
\label{sensorupinf}
\end{figure}

\begin{figure}[b!]
\centering
\includegraphics[scale=0.5]{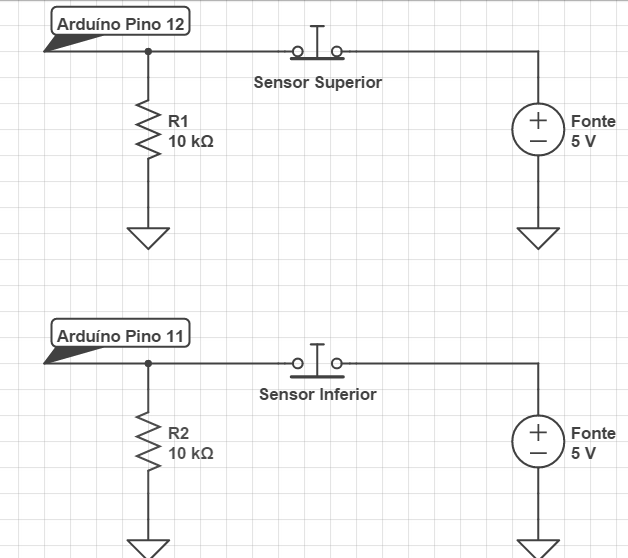}
\caption{ Circuito de apoio. }
\label{circuit1}
\end{figure}

O arranjo experimental \'e mostrado na Figura \ref{esquema}. Uma pequena esfera de a\c co de $9\,$mm de di\^ametro est\'a inicialmente presa entre duas alavancas met\'alicas, uma fixa e outra m\'ovel. Estas duas alavancas fazem parte do sensor superior, veja a Figura \ref{sensorupinf}. Enquanto a esfera de a\c co estiver presa e logo em contato com as alavancas, teremos uma voltagem de $5$V (n\ii vel alto) na porta 12 da placa Ardu\ii no. Quando a alavanca m\'ovel \'e acionada e a esfera liberada, o circuito fica aberto e a voltagem cai para $0$V (n\ii vel baixo) na porta 12. Depois de cair uma altura pr\'e-determinada $h$, a esfera atinge o sensor inferior -- Figura \ref{sensorupinf} -- que consiste em uma base de madeira m\'ovel ligeiramente inclinada que sob o efeito do impacto com a esfera aciona um contato elétrico levando a porta 11 do Arduíno para n\ii vel alto. Assim que a esfera rola para fora da base esta porta retorna ao nível baixo, veja o esquema da Figura \ref{circuit1}. O microcontrolador Arduíno registra o instante da libera\c c\~ao e do impacto, logo, a dura\c c\~ao do tempo queda, depois  permanece inativo durante 10 segundos antes de estar pronto para a pr\'oxima medida. A esfera \'e ent\~ao recolocada na posi\c c\~ao inicial levando o sensor superior novamente para o n\ii vel alto, permitindo uma nova medida. O c\'odigo--fonte utilizado est\'a reproduzido no Ap\^endice. No sensor superior h\'a um furo de 1 mm de di\^ametro onde a esfera deve ser encaixada, garantindo sempre o mesmo ponto de lan\c camento sem risco de mudan\c ca na altura. A altura medida \'e a dist\^ancia entre o ponto de impacto da esfera no sensor inferior e a sua parte inferior, quando esta ainda est\'a presa no sensor superior, assim o di\^ametro da esfera n\~ao causa erros.  A altura $h$ foi medida com uma trena met\'alica com divis\~oes de 1 mm.   O espaço entre os contatos el\'etricos do sensor inferior \'e muito inferior a $1\,$mm.  Conv\'em ressaltar que a esfera met\'alica, assim como os contatos met\'alicos dos sensores, devem estar perfeitamente limpos para evitar interrup\c c\~oes indevidas na comunica\c c\~ao com a placa Ardu\ii no. O c\'odigo--fonte para aquisi\c c\~ao de dados com Arduino \'e mostrado no Ap\^endice. Informa\c c\~oes sobre a programa\c c\~ao da placa Arduino, veja por exemplo, \cite{Salim2011} e refer\^encias ali citadas. 


\section{Resultados experimentais}
Para $h=0,943\,$m com $\delta h= 0,001\,$m e $\delta t =0,001\,$s, os tempos de queda vertical a partir do repouso s\~ao mostrados na Tabela 1. 

\begin{table}[h!]
\centering
\begin{tabular}{l c c c c c c p{1.30cm}}
Med. & $t_i$ (s) & Med. & $t_i$ (s)  & Med. &   $t_i$ (s) \\ 
\hline\hline	
1 & $0,439$ & \ \ 8 & $0,439 $& 15 & $0,439$\\ 
2 & $0,439$ & \ \ 9 & $0,439$& 16 & $0,439$ \\ 
3 & $0,440$ & \ \ 10 & $0,440$ & 17 & $0,439$ \\ 
4 & $0,439$ & \ \ 11 & $0,439$ & 18 &  $0,439$ \\ 
5 & $0,439$ & \ \ 12 & $0,440$ & 19 &  $0,439$ \\ 
6 & $0,439$ & \ \ 13 & $0,440$ & 20 &  $0,439$ \\ 
7 & $0,438$ & \ \ 14 & $0,440$ & 21 &  $0,439$ \\ 
\hline\hline
\end{tabular}
\caption{Medidas dos tempos de queda-livre. A dura\c c\~ao m\'edia da queda \'e de $0,439\,$s.}
\label{tempos}
\end{table}
\vskip 10pt
Os valores obtidos para $g_i$ correspondem ao valor m\'edio de cada medida obtido a partir da equa\c c\~ao (1) da seguinte forma:

\begin{equation}
\bar g_i =\frac{g_{i \,\mbox{\tiny m\'ax}} + g_{i \,\mbox{\tiny m\ii n}}}{2},
\end{equation}
onde

\begin{equation}
g_{i \,\mbox{\tiny m\'ax}} = \frac{2 (h + \delta h)}{(t_i -\delta t)^2};
\end{equation}
e
\begin{equation}
g_{i \,\mbox{\tiny m\ii n}} = \frac{2 (h - \delta h)}{(t_i +\delta t)^2}.
\end{equation}
A incerteza individual $\delta g_i$ pode ser calculada com a express\~ao
\begin{equation}
\delta g_i =\frac{g_{i \,\mbox{\tiny m\'ax}} - g_{i \,\mbox{\tiny m\ii n}}}{2},
\end{equation}
ou do modo usual \cite{Taylor1982}. O resultado \'e $\delta g_i=0,055\,$ m/s$^2$ para todas as vinte e uma medidas individuais. Os resultados para $\bar g_i$ s\~ao mostrados na Tabela 2. O valor m\'edio de $\bar g_i$ \'e $9,778\,$m/s$^2$ e o desvio padr\~ao da m\'edia ou erro padr\~ao \'e $0,012\,$m/s$^2$, logo o experimento nos d\'a

\begin{equation}
g_{\mbox{\tiny exp.}}= 9,78 \pm 0,01 \hskip 2pt \mbox{m/s$^2$ } . 
\end{equation}
\begin{table}[h!]
\centering
\begin{tabular}{c c c c c c c p{1.30cm}}
Med. & $\bar g_i$ (m/s$^2$) & Med. & $\bar g_i$ (m/s$^2$)  & Med. &   $\bar g_i$ (m/s$^2$) \\ 
\hline\hline	
1 & $9,786$ & \ \ 8 & $9,786 $& 15 & $9,786$ \\ 
2 & $9,786$ & \ \ 9 & $9,786 $& 16 & $9,786$ \\ 
3 & $9,742$ & \ \ 10 & $9,742$ & 17 & $9,786$ \\ 
4 & $9,786$ & \ \ 11 & $9,786$ & 18 &  $9,786$ \\ 
5 & $9,786$ & \ \ 12 & $9,742$ & 19 &  $9,786$ \\ 
6 & $9,786$ & \ \ 13 & $9,742$ & 20 &  $9,786$ \\ 
7 & $9,831$ & \ \ 14 & $9,742$ & 21 &  $9,786$ \\ 
\hline\hline
\end{tabular}
\caption{Valores de $\bar g_i$; a incerteza de cada medida individual \'e $\delta g_i=\delta g= 0,055\,$m/s$^2$. }
\label{gmean}
\end{table}
\vskip 12pt
\noindent O resultado dado pela Eq. (6) deve ser comparado com uma aproxima\c c\~ao apropriada ao valor de $g$ local medido por Souza e Santos do Observat\'orio Nacional  \cite{Souza&Santos2010} em 2010, 
\begin{equation}
g_{\mbox{\tiny Rio de Janeiro}}= 978 \, 789, 852 \pm 0,011 \hskip 2pt \mbox{mGal } , 
\end{equation}
onde $1\,\mbox{mGal}= 1\,\times\,10^{-5}\,$m/s$^2$. Para fins de compara\c c\~ao com o resultado experimental obtido basta que consideremos o valor aproximado

\begin{equation}
g_{\mbox{\tiny Rio.}} \approx  9,79 \hskip 2pt \mbox{m/s$^2$ } . 
\end{equation}
O desvio percentual relativo \'e 

\begin{equation}
\frac{|9,79 -9,78|}{9,79}\, \times\, 100 \approx 0,1\,\% .
\end{equation}
\vskip 10pt
Nosso resultado pode tamb\'em ser comparado com os obtidos com outros m\'etodos, veja por exemplo, as refer\^encias \cite{AguiarLaudares2003, Whiteetal2007, DeLuca2011, Schwarz2013}.  \'E poss\ii vel medir tamb\'em com o modelo de placa Arduino que utilizamos (Mega 2560) intervalos de tempo com precis\~ao de microssegundos, mas, no caso do arranjo experimental proposto, os registros das dura\c c\~oes temporais s\~ao inst\'aveis e muitos dados devem ser descartados, o que nos parece inconveniente desde um ponto de vista pedag\'ogico
\section{Conclus\~oes}
Como mencionado anteriormente, por simplicidade, os efeitos de resist\^encia do ar n\~ao s\~ao levados em conta. Desde que $h$ seja uma altura inferior a $1$ metro tais efeitos podem ser desconsiderados em uma primeira abordagem, assim como o retraso na libera\c c\~ao da esfera.  O fator que realmente diferencia a proposta experimental no nível de um laborat\'orio de f\ii sica b\'asica que apresentamos aqui \'e a possibilidade de medir intervalos de tempo com precis\~ao de milissegundos com um arranjo simples e de baixo custo. Observe tamb\'em que embora projetado para o laborat\'orio b\'asico nos n\ii veis m\'edio e universit\'ario, este experimento pode ser realizado facilmente em casa por um aluno ou um amador interessado. 

\section*{Agradecimentos}
Os autores agradecem ao colega Dr. V. Soares pela leitura cr\ii tica do manuscrito original.

\section*{Ap\^endice: c\'odigo de aquisi\c c\~ao de dados com Arduino}
Por raz\~oes pedag\'ogicas, o c\'odigo-fonte utilizado \'e simples e procura obter apenas o tempo de dura\c c\~ao da queda. Para maiores informa\c c\~oes sobre a programa\c c\~ao da placa Arduino ver \cite{Salim2011}.
\vskip 10pt
\begin{verbatim}
int sensorSUP = 12;
int sensorINF = 11;
int estadoSUP;
unsigned long startTime;
unsigned long stopTime;
void setup() {
  Serial.begin(9600);
  pinMode (sensorSUP, INPUT);
  pinMode (sensorINF, INPUT);
  estadoSUP = 1 ;
  Serial.println("Pronto para medir o tempo de queda");
  Serial.println("                            ");
}
void loop()
{
  if (digitalRead(sensorSUP) == LOW && (digitalRead(sensorINF)) == LOW && estadoSUP == 1)
  {
    startTime = millis();
    Serial.print(" Inicio da medida ---");
    Serial.print("Esfera caindo...Contanto tempo...");
    estadoSUP = 0;
  }
  if (digitalRead(sensorINF) == HIGH && (digitalRead(sensorSUP)) == LOW)
  {
    estadoSUP = 1;
    stopTime = ( millis() - startTime);
    Serial.print("Tempo de queda = ");
    Serial.print(stopTime);
    Serial.print(" milisegundos (ms)| Aguarde 10 segundos...");
    delay(10000);
    Serial.println("Pronto para proxima medida");
  }
}
\end{verbatim}
\end{document}